\documentclass[12pt]{article}
\usepackage{amssymb}
\usepackage{graphicx}
\usepackage{amsmath}

\newcommand{\be}{\begin{eqnarray}}
\newcommand{\ee}{\end{eqnarray}}
\newcommand{\ba}{\begin{array}}
\newcommand{\ea}{\end{array}}

\newcommand{\chil}[1]{\stackrel{o}{#1}}

\newcommand{\nn}{\nonumber}
\renewcommand{\(}{\left(}
\renewcommand{\)}{\right)}
\renewcommand{\[}{\left[}
\renewcommand{\]}{\right]}

\begin{document}

\rightline{\textsl{\date{\today}}} \vspace{0.5cm}
\begin{center}
{\Large Leading infrared logarithms for  $\sigma$-model with 
fields on arbitrary Riemann manifold\footnote{Devoted to memory of Alexandre Nikolaevich Vasiliev.}}\\
\vspace{0.35cm}
 Maxim V. Polyakov$^{1,2}$ and Alexei  A. Vladimirov$^{2}$\\

\vspace{0.35cm}
$^1$Petersburg Nuclear Physics
Institute, Gatchina, 188300, Russia\\
$^2$Institut f\"ur Theoretische Physik II, Ruhr--Universit\"at
Bochum, D--44780 Bochum, Germany
\end{center}

\begin{abstract}
\noindent
We derive non-linear recursion equation for the leading infrared logarithms (LL) in  
four dimensional $\sigma$-model with fields on an arbitrary Riemann manifold. 
The derived equation allows one to compute leading infrared logarithms to essentially unlimited loop order
in terms of geometric characteristics of the Riemann manifold.\\
\vspace{-10pt}

\noindent
 We reduce the solution of the $SU(\infty)$ principal chiral field in arbitrary number of dimensions 
in the LL approximation to the solution of very simple recursive equation.
This result paves a way to the solution of the model in arbitrary number of dimensions
at $N\to\infty$.
\end{abstract}

%\PACS{13.60.Le\and14.20.Gk}

\section{Introduction}
Effective Field Theories (EFTs) are {\it non-renormalizable} field
theories, which allow the investigation of the infrared (low-energy) behaviour of
various physical systems (see different examples in Refs.~\cite{ANV,Weinberg}).
The standard tool for the studies of the asymptotic behaviour of {\it renormalizable}
field theories is the method of renormalization group equations (RGEs).
In the case of EFTs the method of RGEs must be modified  as the number of counterterms
increases rapidly with the loop order. For subtleties of the renormalization in quantum field theory
see an excellent book by Alexandre Nikolaevich Vasiliev \cite{ANV} . 

A possibility of the systematic construction of RGEs for
non-renormalizable quantum field theories was demonstrated  in
Ref.~\cite{BC_2003}. In particular, it was shown that the series
of the leading logarithms (LLs) can be obtained by calculation of
one loop diagrams. However, the solution of the RGEs derived in
Ref.~\cite{BC_2003}  requires the calculation of non-trivial
one-loop diagrams. The number of the diagrams increases rapidly with
the loop order. Therefore, the implementation of this method in
practice is not an easy task. The method of Ref.~\cite{BC_2003}
has been applied  in Ref.~\cite{bijnens} for the calculation of LLs
up to five-loops for the decay constant and pion mass,
 and up to four-loops for meson-meson scattering in the massive $O(N+1)/O(N)$
sigma model. In Ref.~\cite{Biss2} the authors, using dispersive
methods, calculated the three-loop LLs to $\pi\pi$ scattering in
massless Chiral Perturbation Theory (ChPT).

Recently, a completely different method for the
calculation of LLs in a wide class
of non-renormalizable {\it massless} field theories was developed in Refs.~\cite{MKV_LLog,MKV_FF,KPV}.
The { non-linear recursion} equations derived in Refs.~\cite{MKV_LLog,MKV_FF,KPV} allow one to obtain the LLs contributions
to practically unlimited loop order without performing non-trivial loop calculations at each loop order.

In the present paper we compute the leading infrared logs contribution for the four dimensional $\sigma$-model
with fields on arbitrary Riemann manifold.  One of our main aim here is to 
obtain the interpretation of LLs in terms of geometric characteristics of the Riemann manifold.   
In particular, this approach allows us to reduce the solution of the $SU(\infty)$ principal chiral field in arbitrary 
number of dimensions 
in the LL approximation to the solution of very simple recursive equation.
This result paves a way to the solution of the model in arbitrary number of dimensions
at $N\to\infty$.

The paper is organized as follows.
 The most general main  results of this paper are given in Eqs.~(\ref{initial},\ref{main}) of Section~\ref{general}. 
 We also give in Section~\ref{general}, it seems for the first time, 2-loop results for general $\sigma$-model, see Eq.~(\ref{twoloop}).   
The way how the general method of Section~\ref{general} works is demonstrated on the ``routine example" of $N$ dimensional
 sphere as a target space in Section~\ref{SN}.
The most interesting and non-trivial results for $SU(\infty)$
principal chiral field are given by Eqs.~(\ref{sunLN},\ref{quint}) of Section~\ref{suoo}.

\section{The $\sigma$-model with fields on arbitrary Riemann manifold}
\label{general}

We consider the most general $\sigma$-model with the fields on an arbitrary Riemann
manifold. For two space-time dimensions the corresponding model is renormalizable
\cite{Friedan} and LL infrared asymptotic is determined by the one-loop RGE, which is just the Ricci flow equation for the metric.
In higher space-time dimensions the calculation of the LLs is highly non-trivial
task, to our best knowledge only one loop LLs in four dimensions were computed
for a general $\sigma$-model \cite{honerkamp,Howe}.  Here we show how to compute the LLs to an arbitrary loop order for
a general  $\sigma$ model  in four space-time dimensions\footnote{The generalization to $D\geq 4$ is simple and
can be done using method of Ref.~\cite{KPV}.} .
 
The action is given by the following expression:

\be
S=\int d^4 x\ \frac 12\  g_{ab}(\phi)\ \partial_\mu \phi^a\partial_\mu\phi^b\, ,
\label{action}
\ee
where $g_{ab}(\phi)$ is a metric on a Riemann manifold. Without loss of generality we consider here
a compact Riemann manifold with positive signature of the metric.
Using the freedom of the coordinates ($\phi^a$) choice on the Riemann manifold we  fix the metric such
that $g_{ab}(0)=\delta_{ab}$ and $g_{ab}(\phi)\phi^b=\delta_{ab}\phi^b$. The latter condition corresponds
to the choice of the normal coordinates on the Riemann manifold \cite{Riemann}. In these coordinates
the geodesics in the vicinity of $\phi^a=0$ are simple straight lines and the metric has the following
expansion around $\phi^a=0$:

\be
\label{gexp}
g_{ab}(\phi)=\delta_{ab}-\frac{1}{3}\ \chil{R}_{acbd}\phi^c\phi^d+O(\phi^3)\, ,
\ee
where $\chil{R}_{acbd}$ is the Riemann curvature tensor at $\phi^a=0$. Substituting the expansion (\ref{gexp})
in the action (\ref{action}) we can easily compute the tree level scattering amplitude of $\phi^a+\phi^b\to \phi^c+\phi^d$
with the result:

\be
\label{amptree}
A_{abcd}^{\rm tree}(s,\cos\theta)=\frac{s}{2}\left(-\left[\chil R_{acbd}+\chil R_{adbc}\right] P_0(\cos\theta)+
\chil R_{abcd} P_1(\cos\theta)\right).
\ee 
Here $s$ is the Mandelstam variable, $\theta$ is the centre of mass scattering angle and $P_l$ are Legendre
polynomials.  This expression provides the leading infrared asymptotic of the scattering amplitude.
The corrections to this result arise from the loop contributions. Simple power counting arguments
show that the $(n-1)$-loop contribution has the following form of the LLs contribution $s [s\ln(\mu^2/s)]^{n-1}$.
For the $l$'s partial wave amplitude defined as:
\be
t^l_{abcd}(s)&=&\frac{1}{64\pi}\ 
\int_{0}^\pi d\theta\ \sin\theta\ A_{abcd}(s,\cos\theta)P_l(\cos\theta),
\ee
we can write the general form of the LL expansion:
\begin{eqnarray}\label{t^I_l}
 t^l_{abcd}(s)=\frac{\pi}{2}\frac{1}{2 l+1}\sum_{n=1}^\infty (\omega_{nl})_{abcd} \left(\frac{ s}{16 \pi^2}\right)^n
\ln^{n-1}\Big(\frac{\mu^2}{s}\Big) 
+\mathcal{O}(\text{NLL}).
\end{eqnarray}
Here $(\omega_{nl})_{abcd}$ are the LLs coefficients which depend only on the geometric characteristics
of the target space. These coefficients are non-zero only for $l\leq n$, where $(n-1)$ is the number of loops.
The tree level amplitude (\ref{amptree}) corresponds to $n=1$, in this case only
$l=0$ and $l=1$ coefficients are non-zero:
\be
\label{initial}
(\omega_{10})_{abcd}&=&-\frac 12 \left(\chil R_{acbd}+\chil R_{adbc}\right),\\
\nonumber
(\omega_{11})_{abcd}&=&\frac 12 \ \chil R_{abcd}.
\ee

Now we can use the general method of Ref.~\cite{KPV} to obtain the recursion equation which
allows us to express the higher loop LL coefficients in terms of the tree level ones (\ref{initial}).
The method of Ref.~\cite{KPV} is based on the general principles of a quantum field theory --
analyticity, unitarity and crossing. We refer the reader to \cite{KPV} for the description of the general method, here we present
the final result for the recursion equation:

\begin{eqnarray}
\label{main}
&&\(\omega_{nl}\)_{abcd}=\frac{1}{2(n-1)}\[\sum_{i=1}^{n-1} \frac{\(\omega_{il}\)_{ab\alpha\beta}\(\omega_{n-i,l}\)^{\beta\alpha}_{~~~cd}}{2l+1}\right.\\ &&\nn \left.+\sum_{i=1}^{n-1}\sum_{l'=0}^{n-1}\frac{
\(\omega_{il'}\)_{ad\alpha\beta}\(\omega_{n-i,l'}\)^{\beta\alpha}_{~~~cb}\Omega_{n}^{l'l}+\(-1\)^{l+l'}
\(\omega_{il'}\)_{ac\alpha\beta}\(\omega_{n-i,l'}\)^{\beta\alpha}_{~~~bd}\Omega_n^{l'l}}{2l'+1}\].
\end{eqnarray}
Here $\Omega_n^{l'l}$ is the crossing matrix in the angular momentum space, it has the following form:

\be
\Omega_n^{l'l}=\frac{2 l+1}{2^{n+1}}\int_{-1}^1 dz\ P_{l'}\left(\frac{z+3}{z-1}\right) P_l(z) (z-1)^n.
\ee
Eq.~(\ref{main}) expresses the higher loop LL coefficients in terms of the lower loop. The starting point for the
recursion is given by the tree level expressions (\ref{initial}).

One can easily check that the solution of the equation (\ref{main}) satisfies the following symmetries:
\begin{eqnarray}
(\omega_{nl})_{abcd}=(-1)^l(\omega_{nl})_{bacd}=(-1)^l(\omega_{nl})_{abdc}=(\omega_{nl})_{cdab}.
\end{eqnarray}
These symmetries are actually the consequence of the Bose symmetry and the parity conservation.
Additionally Eq.~(\ref{main}) posses the following symmetries:
\begin{eqnarray}\label{w_sym}
\(\sum_{l'=0}^n\omega_{nl'}\Omega_n^{l'l}\)_{abcd}=(\omega_{nl})_{adcb},~~~\(\sum_{l'=0}^n\omega_{nl'}(-1)^{l+l'}\Omega_n^{l'l}\)_{abcd}=(\omega_{nl})_{acbd},
\end{eqnarray}
which might be very useful for possible exact solution of the recursion equation (\ref{main}).

We did not find yet the analytic solution of the equation (\ref{main}), therefore we give here expressions for the
LL coefficients to the 2-loop order. The 1-loop LL coefficients in terms of geometric 
characteristics have the following form:
\begin{eqnarray}
\nn
&&\(\omega_{20}\)_{abcd}=%\frac{1}{2}R_{a\beta_1b\beta_2}R_{c~~d}^{~\beta_1~\beta_2}-\frac{1}{8}R_{ab\beta_1\beta_2}R^{\beta_1\beta_2}_{~~~~cd} \\ \nn
\frac{1}{4}\[R_{a\beta_1b\beta_2}R_{c~~d}^{~\beta_1~\beta_2}+R_{a\beta_1b\beta_2}R_{c~~d}^{~\beta_2~\beta_1}\]\\ \nn
&&+\frac{1}{6}\[R_{a\beta_1c\beta_2}R_{b~~d}^{~\beta_1~\beta_2}+R_{a\beta_1d\beta_2}R_{b~~c}^{~\beta_1~\beta_2}\] -\frac{5}{72}\[R_{ac\beta_1\beta_2}R^{\beta_1\beta_2}_{~~~~bd}+R_{ad\beta_1\beta_2}R^{\beta_1\beta_2}_{~~~~bc}\],\\
\label{oneloop}
&&\(\omega_{21}\)_{abcd}=-\frac{1}{24}R_{ab\beta_1\beta_2}R^{\beta_1\beta_2}_{~~~~cd} \\ 
\nn
 &&+\frac{1}{4}\[R_{a\beta_1c\beta_2}R_{b~~d}^{~\beta_1~\beta_2}-R_{a\beta_1d\beta_2}R_{b~~c}^{~\beta_1~\beta_2}\] -\frac{1}{12}\[R_{ac\beta_1\beta_2}R^{\beta_1\beta_2}_{~~~~bd}-R_{ad\beta_1\beta_2}R^{\beta_1\beta_2}_{~~~~bc}\],\\
\nn
&&\(\omega_{22}\)_{abcd}=\frac{1}{12}\[R_{a\beta_1c\beta_2}R_{b~~d}^{~\beta_1~\beta_2}+R_{a\beta_1d\beta_2}R_{b~~c}^{~\beta_1~\beta_2}\] \\
\nn
&&~~~~~~~
-\frac{1}{72}\[R_{ac\beta_1\beta_2}R^{\beta_1\beta_2}_{~~~~bd}+R_{ad\beta_1\beta_2}R^{\beta_1\beta_2}_{~~~~bc}\].
\end{eqnarray}
Here and below all Riemann tensors are at $\phi^a=0$.
The 2-loop result is the following:
\begin{eqnarray}
\label{twoloop}
\(\omega_{30}\)_{abcd}&=&\chi_{abcd}-\frac12 \psi_{abcd}+\frac{1}{3}\psi_{adbc}-\frac14 \chi_{adbc},\\
\nn
\(\omega_{31}\)_{abcd}&=&\frac12 \psi_{abcd}-\frac{1}{2}\psi_{adbc}-\frac{9}{20}\chi_{adbc},\\
\nn
\(\omega_{32}\)_{abcd}&=&\frac{1}{6} \psi_{adbc}-\frac14\chi_{adbc},\\
\nn
\(\omega_{33}\)_{abcd}&=&\frac{1}{20}\chi_{adbc},
\end{eqnarray}
where 
\begin{eqnarray*}
\chi_{abcd}&=&\[R_{a\beta_1b\beta_2}\(-\frac{1}{72}R_{c\beta_3~~\beta_4}^{~~~\beta_1}R_d^{\beta_3\beta_2\beta_4} -\frac{7}{144}R_{c\beta_3~~\beta_4}^{~~~\beta_1}R_d^{\beta_4\beta_2\beta_3}\right.\right.
\\ &&~~~~~~~~~~~~~~~~\left.-\frac{1}{48}R_{c\beta_3~~\beta_4}^{~~~\beta_2}R_d^{\beta_4\beta_1\beta_3}\)  +(a\rightarrow b\rightarrow d \rightarrow c\circlearrowleft)
\\
&& + \frac{1}{144}R_{a\beta_1 b\beta_2}\(R_{c\beta_3 d\beta_4}-R_{c\beta_4 d\beta_3}\)R^{\beta_1\beta_2\beta_3\beta_4} \\ && \left.- \frac{1}{8}R_{a\beta_1 b\beta_2} \(R_{c\beta_3 d \beta_4}+R_{c\beta_4 d \beta_2}\)R^{\beta_1\beta_3\beta_2\beta_4}\] \\
&& + \(c\leftrightarrow b\),
\end{eqnarray*}
\begin{eqnarray*}
\psi_{abcd}&=&\[R_{ab\beta_1\beta_2}\(-\frac{1}{72}R_{c\beta_3~~\beta_4}^{~~~\beta_1}R_d^{~\beta_3\beta_2\beta_4} -\frac{1}{36}R_{c\beta_3~~\beta_4}^{~~~\beta_1}R_d^{~\beta_4\beta_2\beta_3}\)\right. \\ &&\left.
~~~~+\(\begin{array}{c}a\leftrightarrow c \\ b\leftrightarrow d\end{array}\)- \(b\leftrightarrow c\)-(a\rightarrow b\rightarrow d \rightarrow c\circlearrowleft)\] \\ && +\[R_{b\beta_1 d\beta_2}\(\frac{1}{36}R_{a\beta_3~~\beta_4}^{~~~\beta_2}R_{c}^{~\beta_3\beta_1\beta_4} +\frac{1}{72}R_{a\beta_3~~\beta_4}^{~~~\beta_1}R_{c}^{~\beta_3\beta_2\beta_4} \right.\right. \\ && \left.
~~+\frac{17}{144}R_{a\beta_3~~\beta_4}^{~~~\beta_2}R_{c}^{~\beta_4\beta_2\beta_3}+\frac{13}{144}R_{a\beta_3~~\beta_4}^{~~~\beta_1}R_{c}^{~\beta_4\beta_2\beta_3}\) \\&& \left. +\(\begin{array}{c}a\leftrightarrow b \\ c\leftrightarrow d\end{array}\)\]+\frac{3}{16}R_{a\beta_1 c\beta_2}R_{b\beta_3 d\beta_4}R^{\beta_1\beta_3\beta_2\beta_4}
\\&& +\(\frac{1}{144}R_{ab\beta_1\beta_2}R_{cd\beta_3\beta_4}-\frac{1}{72}R_{a\beta_1 c\beta_2}R_{b\beta_3 d\beta_4}\)R^{\beta_1\beta_2\beta_3\beta_4}
\end{eqnarray*}

The expressions for higher loops can be easily obtained by further iterations of Eq.~(\ref{main}). We do not give them, as
the corresponding expressions are rather busy. Instead, we discuss below the particular cases of the Riemann
manifolds: the $N$-dimensional sphere $S^N$ and the group manifold corresponding to a simple Lie group
$G$. The latter case is the principal chiral field $\sigma$-model that has much in common with Yang-Mills theory.

\section{The $\sigma$-model  with fields on a sphere $S^N$}
\label{SN}

The metric on the $N$-dimensional sphere $S^N$ in the normal coordinates has the following form:
\be
\label{sNmetric}
g_{ab}(\phi)=F^2\(\delta_{ab}\ \frac{\sin^2\(\frac{|\phi|}{F}\)}{|\phi|^2}+\frac{\phi_a\phi_b}{|\phi|^2} \[
\frac{1}{F^2}-\frac{\sin^2\(\frac{|\phi|}{F}\)}{|\phi|^2}
\]\),
\ee
where $|\phi|\equiv \sqrt{\phi_a\phi^a}$ and $F$ is the radius of $S^N$. For the theory (\ref{action}) in
four space-time dimensions the constant $F$ has the dimension of mass, and for $S^3$ corresponds to the
pion decay constant of the effective chiral Lagrangian of two flavour QCD. 
The Riemann and Ricci tensors, and scalar curvature for $S^N$
have the following
form:
\be
R_{abcd}=\frac{1}{F^2}\(g_{ac}g_{bd}-g_{ad}g_{bc}\),\ \ \ R_{ab}=\frac{N-1}{F^2}\ g_{ab},\ \ \ R=\frac{N(N-1)}{F^2}.
\ee
Substituting these expressions to the general solutions (\ref{oneloop},\ref{twoloop}) one obtains:
\be
\(\omega_{20}\)_{abcd}&=&
\frac{1}{18 F^4}\[(3N-1) (\delta_{ad}\delta_{bc}+\delta_{ac}\delta_{bd})+(9 N-17)\delta_{ab}\delta_{cd} \],\\
\nn
\(\omega_{21}\)_{abcd}&=&\frac{1}{4 F^4} (N-3) (\delta_{ad}\delta_{bc}-\delta_{ac}\delta_{bd}),\\
\nn
\(\omega_{22}\)_{abcd}&=&
\frac{1}{36 F^4}\[(3N-5) (\delta_{ad}\delta_{bc}+\delta_{ac}\delta_{bd})+4\delta_{ab}\delta_{cd} \],
\ee
for 1-loop LL coefficients. The 2-loop coefficients have the following form:
\be
\(\omega_{3l}\)_{abcd}&=&
\frac{1}{288 F^6}\[a_{3l}\  (\delta_{ad}\delta_{bc}+\delta_{ac}\delta_{bd})+b_{3l}\ \delta_{ab}\delta_{cd} \],\ {\rm
for}\ l=0,2\\
\nn
\(\omega_{3l}\)_{abcd}&=&\frac{1}{1440 F^6} c_{3l} \ (\delta_{ad}\delta_{bc}-\delta_{ac}\delta_{bd}),,\ {\rm
for}\ l=1,3
\ee
with
\be
a_{30}&=&-59 + 27 N - 18 N^2,\ b_{30}=86 - 158 N + 72 N^2,\\
\nn
a_{32}&=&-39 + 47 N - 18 N^2,\ b_{32}=10+10 N,\\
\nn
c_{31}&=&441 - 333 N + 162 N^2,\ c_{33}=49 - 37 N + 18 N^2.
\ee
In the above results for the first two loops one can note a systematic pattern for the 
leading large $N$ coefficients. Indeed, one can easily derive\footnote{One has to use the
isometries of $S^N$ manifold, see next Section. } the large $N$ solution
of Eq.~(\ref{main}) for the case of $S^N$ \cite{MKV_LLog,KPV}. The corresponding solution has the form:
\be
\nn
\(\omega_{nl}\)_{abcd}^{{\rm Large}\ N}&=&
\frac{1}{F^2} \(-\frac{N}{2 F^2}\)^{n-1}
\Big [ \frac{ (2 l+1)n\ n!}{(n+l+1)!}\(\delta_{ad}\delta_{bc}+(-1)^l \delta_{ac}\delta_{bd}\) \\
\label{largeNsolution}
&+& (-1)^{n-1} \delta_{l0}\ \delta_{ab}\delta_{cd} \Big].
\ee
We see that the $\sigma$-model with fields on $S^N$ can be solved in the large $N$-limit
for arbitrary number of space-time dimensions. It is well known result. The record calculations of
the $1/N$ corrections were performed by A.N.~Vasiliev et al. \cite{ANVN} by very elegant method of the
conformal bootstrap.
From the large $N$ solution  (\ref{largeNsolution}) we see that the amplitude, after summation
of LL's, has a pole for $l=0$ and $O(N)$ singlet channel. This pole corresponds to 
the contribution of the auxiliary scalar field which one introduces to solve the sigma model on $S^N$
in the large $N$ limit.

We note that all results obtained in this section for $S^N$ are valid also for any 
simple connected conformally flat Riemann manifold.

\section{Principal chiral field,  $SU(\infty)$ case especially}
\label{suoo}
An important case is the $\sigma$-model with the target space $\frac{G\times G}{G}$, where $G$ is a simple
Lie group. This is the model of
principal chiral field.  The metric in normal coordinates on $\frac{G\times G}{G}$ has the form:
\be
g_{ab}(\phi)=\int_0^1 d\alpha\ (1-\alpha) {\rm tr}\(\exp\(i\alpha\frac{ t^c\phi^c}{F}\)\ t^a\ \exp\(-i\alpha\frac{  t^c\phi^c}{F}\)\ t^b \),
\ee
where $t^a$ are the generators of the group $G$ in the fundamental representation 
which are normalized by ${\rm tr}(t^a t^b)=2 \delta^{ab}$, $F$ is the parameter of mass dimension one. 
For $G=SU(N)$ this parameter corresponds to Nambu-Goldstone boson decay constant in the chiral Lagrangian 
for QCD with $N$ massless quark flavours. For what follows we consider the case of $G=SU(N)$.
The $SU(N)$ principal chiral field model is many respects is similar to the Yang-Mills theory, for example, its large $N$
limit corresponds to the summation of the planar diagrams. Despite many hopes \cite{AMP} this model is not
solved in the large $N$ limit for arbitrary number of space-time dimensions. We shall discuss this limit in LL
approximation below. 

The Riemann and Ricci tensors, and the scalar curvature for $SU(N)\times SU(N)/SU(N)$ manifold
have the following form:
\be
R_{abcd}=\frac{1}{8F^2}\ {\rm tr}\([t^a,t^b] [t^c,t^d]\)+O(\phi^2),\ \ R_{ab}=\frac{N}{F^2}\ g_{ab}, \ \ R=\frac{N(N^2-1)}{F^2}.
\ee 
In principle, one can use these expressions for the our main equations (\ref{main},\ref{initial})  to obtain
LL coefficients. However, the results, especially for high loop orders, are rather cumbersome. Therefore,
at this point it is wise to use the isometries of the $SU(N)\times SU(N)/SU(N)$ manifold. For the case of 
the amplitude   (\ref{t^I_l}) the isometries allow us to decompose the amplitude in the projectors
onto the irreducible representations of the $SU(N)$ group:
\be
t^l_{abcd}(s)=\sum_{R=1}^7 P_{abcd}^R\ t_R^l(s),
\ee
where $P_{abcd}^R$ are projectors onto the irreducible representations of the $SU(N)$ group
which arise in the product of Adj$\times$Adj. Generically, these are seven  representations with the dimensions
\footnote{For the reader experienced in $SU(3)$ these are $d_R=\(1,8_S,-,27,8_A,10, \overline{10}\)$.}:
\be
\label{reps}
d_R&=&\( 1, N^2-1, \frac{N^2(N+1)(N-3)}{4},\frac{N^2(N-1)(N+3)}{4},\right.\\
\nn
&&\left. N^2-1,\frac{(N^2-4)(N^2-1)}{4},\frac{(N^2-4)(N^2-1)}{4}\).
\ee 
The derivation and the explicit expressions for the projectors $P_{abcd}^R$ can be found in \cite{lexadisser}. 
We note that the representations with numbers $R=2$ and $R=5$ correspond to the symmetric and the antisymmetric
adjoint representations of the $SU(N)$ group. In Ref.~\cite{lexadisser} it was shown that in the large $N$
limit only these two adjoint representations survive, the contribution of other representation is the $1/N$ correction.
That is very interesting observation, because it allows one to reduce considerably the number of geometric objects in the
large $N$ limit. 

The solution for the LL coefficients to {\it arbitrary loop order}  at $N\to \infty$ for 
 the $SU(N)$ principal chiral field has the following form:

\be
%\nn
%(\omega_{nl})_{abcd}^{{\rm Large}\ N}&=&\frac{\rho_{nl}}{16 F^2} \  
%\(\frac{N}{2 F^{2}}\)^{n-1}\({\rm tr}(\{t^a,t^b\} \{t^c,t^d\})-\frac{1}{N} \delta_{ab}\delta_{cd} \),\ \
%{\rm for}\ {\rm even}\ l, \\
\label{sunLN}
(\omega_{nl})_{abcd}^{{\rm Large}\ N}&=&\frac{\rho_{nl}}{16 F^2} \  \(\frac{N}{ 2 F^{2}}\)^{n-1}\  {\rm tr}([t^a,t^b] [t^c,t^d]) ,\ \ 
{\rm for}\ {\rm odd}\ l,
\ee
where the coefficients $\rho_{nl}$ satisfy simple recursion equation:
\be
\label{quint}
\rho_{nl}=\frac{1}{2(n-1)} \sum_{i=1}^{n-1}\sum_{l'=0}^n \frac{\rho_{il'}\ \rho_{n-i, l'}}{2 l'+1} \(\delta^{l'l}+\Omega_n^{l'l}\),
\ee
with the initial conditions $\rho_{10}=\rho_{11}=1$. The expression for even $l$ is rather complicated
and we do not have place to show it. It can be easily obtained from Eq.~(\ref{sunLN}) with help of relations
(\ref{w_sym}). In any case, the amplitude for even $l$ is expressed in terms of the same coefficients
$\rho_{nl}$ (\ref{quint}).

We reduced the solution of the $SU(\infty)$ principal chiral field in arbitrary number of dimensions 
in the LL approximation to the solution of very simple recursive equation (\ref{quint}). It is remarkable
equation, it possesses rich symmetries. 
Unfortunately we did not find yet its analytical solution\footnote{As a side remark we note that 
Eq.~(\ref{quint}) without the term $\sim \Omega_n^{l'l}$ has very simple solution
$\rho_{nl}=\frac{1}{2^{n-1}}\ \delta_{l0}+\frac{1}{6^{n-1}}\ \delta_{l1}$. }, however
it can be solved numerically to practically unlimited loop order. Studies of this equation we shall present 
elsewhere.

\section{Conclusions and outlook}
We derived the non-linear recursion relation (\ref{main}) for the LL coefficients in the four dimensional sigma
model with fields on arbitrary Riemann manifold. Being supplemented by the initial conditions (\ref{initial})
this equation allows one to obtain the LL coefficients in terms of the geometric characteristics of the Riemann
manifold. We calculated explicitly two loop LLs for arbitrary $\sigma$-model.
We can speculate that 
it might be possible to obtain Eq.~(\ref{main}) as a some kind of equations of motion for a non-local
object in the Riemann manifold. If one would manage this, the LL coefficient could be
obtained by the expansion of the object in its non-locality.

Eq.~(\ref{main}) being applied to 4D $SU(N)$ principal chiral field allowed us to reduce 
the solution of this theory at $N\to\infty$ and LL approximation to the solution 
of simple and nice equation (\ref{quint}). We hope that this equation can be solved analytically
with help of rich symmetries it possesses. Such solution can provide a clue for 
the spectrum of  $SU(\infty)$ principal chiral field in arbitrary number of space-time dimensions.

\section*{Acknowledgements}
We are thankful to D.~Diakonov for encouraging us to make the calculations presented here.
Many illuminating discussions with Dmitri Diakonov,  Kolya Kivel, Julia Koschinski and Victor Petrov
were big help for us.
The work is supported in parts by German Ministry for Education and Research (grant 06BO9012) and by Russian Federal Programme ``Research and Teaching Experts in Innovative Russia"
(contract 02.740.11.5154).

%%%%%%%%%%%%%%%%%%%%%%%%%%%%%%%%%%%%%%%%%%%%%%%%%%%%%%

\end{document}